\documentclass[twocolumn,prb,showpacs,floatfix]{revtex4}

\usepackage{graphicx}
\usepackage{bm}
\usepackage{color}
\usepackage{amsmath}
\usepackage{amsfonts}
\usepackage{amssymb}
\usepackage{epstopdf}

\begin{document}

\preprint{APS/123-QED}

\title{Upper Critical Field of Pressure-Induced Superconductor EuFe$_2$As$_2$}

\author{Nobuyuki Kurita$^{1,2}$} \author{Motoi Kimata$^{1,2}$} \author{Kota Kodama$^{1,3}$} \author{Atsushi Harada$^1$} \author{Megumi Tomita$^1$} 
\author{Hiroyuki S. Suzuki$^1$}  \author{Takehiko Matsumoto$^1$} \author{Keizo Murata$^4$} \author{Shinya Uji$^{1,2,3}$} \author{Taichi Terashima$^{1,2}$}
\affiliation{$^1$National Institute for Materials Science, Tsukuba, Ibaraki 305-0003, Japan \\
$^2$JST, Transformative Research-Project on Iron Pnictides (TRIP), Chiyoda, Tokyo 102-0075, Japan \\
$^3$Graduate School of Pure and Applied Sciences, University of Tsukuba, Ibaraki 305-0003, Japan \\
$^4$Division of Molecular Materials Science, Graduate School of Science, Osaka City University, Osaka 558-8585, Japan}

\date{\today}

\begin{abstract}

We have carried out high-field resistivity measurements up to 27\,T in EuFe$_2$As$_2$ at $P$\,=\,2.5\,GPa, a virtually optimal pressure for the $P$-induced superconductivity, where $T_\mathrm{c}$\,=\,30\,K.  
The $B_\mathrm{c2}$$-$$T_\mathrm{c}$ phase diagram has been constructed in a wide temperature range with a minimum temperature of 1.6\,K ($\approx$\,0.05\,$\times$\,$T_\mathrm{c}$), for both $B$\,$\parallel$\,$ab$\,($B_\mathrm{c2}^\mathrm{ab}$) and $B$\,$\parallel$\,$c$\,($B_\mathrm{c2}^\mathrm{c}$).  
The upper critical fields $B_\mathrm{c2}^\mathrm{ab}$(0) and $B_\mathrm{c2}^\mathrm{c}$(0), determined by the onset of resistive transitions, are 25\,T and 22\,T, respectively, which are significantly smaller than those of other Fe-based superconductors with similar values of $T_\mathrm{c}$.
The small $B_\mathrm{c2}(0)$ values and the $B_\mathrm{c2}(T)$ curves with positive curvature around 20\,K can be explained by a multiple pair-breaking model that includes the exchange field due to the magnetic Eu$^{2+}$ moments.  
The anisotropy parameter, $\Gamma$\,=\,$B_\mathrm{c2}^{ab}/B_\mathrm{c2}^{c}$, in EuFe$_2$As$_2$ at low temperatures is comparable to that of other ``122" Fe-based systems.

\end{abstract}


\maketitle


The discovery of superconductivity in LaFeAs(O,F) at $T_\mathrm{c}$\,=\,26\,K\,\cite{Kamihara} has inspired experimental and theoretical research on a group of FeAs-layered superconductors (SCs).\cite{review}
Basically, Fe-based high-$T_\mathrm{c}$ superconductivity\,\cite{Kito2008,ZARen2008a,Wang_56K} occurs when the antiferromagnetic (AF) order in the mother compounds is suppressed
by means of carrier doping,\cite{Kamihara} application of pressure ($P$),\cite{Alireza} or isovalent substitution.\cite{Ren_EuFe2AsP2}
As compared to other methods in studying such interplay between magnetism and superconductivity, pressure experiments have a significant advantage in that they are free from random impurity potentials that may distort the underlying physics of the low-lying energy states.
Among the various crystal structures, tetragonal ThCr$_2$Si$_2$-type (``122") compounds have been investigated more intensively owing to the availability of highly-pure stoichiometric single crystals. 
In particular, $A$Fe$_2$As$_2$ ($A$\,=\,Sr, Eu) exhibits $P$-induced bulk superconductivity with $T_\mathrm{c}$ of order 30\,K.\cite{Alireza, Matsubayashi_Sr, Terashima_Eu1}
In contrast, superconductivity under hydrostatic pressure is not exhibited by CaFe$_2$As$_2$,\cite{Yu_Helium} and its occurrence in BaFe$_2$As$_2$ has not been established definitively.\cite{Matsubayashi_Sr,Yamazaki_Ba}

A fundamental characteristic of SCs is the upper critical field $B_\mathrm{c2}$. 
$B_\mathrm{c2}$ has its roots in the breakdown of Cooper pairs; hence, the $B_\mathrm{c2}$$-$$T_\mathrm{c}$ phase diagram provides important insights into the pairing mechanism of high-$T_\mathrm{c}$ superconductivity. 
Thus far, to our knowledge, there has been no reports on $B_\mathrm{c2}$ for $P$-induced Fe-based SCs at low temperatures.
This is mainly attributed to the difficulty in conducting high-pressure experiments on high-$T_\mathrm{c}$ SCs under a high-field. 
In the case of SrFe$_2$As$_2$ ($T_\mathrm{c}$\,=\,30\,K at 4.2\,GPa), a field of 8\,T brings about a small reduction in $T_\mathrm{c}$ (i.e., to 27\,K) for $B$\,$\parallel$\,$ab$.\cite{Kotegawa_Sr}
Assuming an orbitally limited case,\cite{WHH1966} $B_\mathrm{c2}$ ($T$\,=\,0\,K) could exceed 60\,T.\cite{Kotegawa_Sr}
However, the low-temperature region of the $B_\mathrm{c2}$ curve, where paramagnetic and/or multiband effects may play important roles,\cite{twobandLaFeAsO1} has not been investigated.

In the case of EuFe$_2$As$_2$ ($T_\mathrm{c}$\,=\,30\,K at $\sim$\,2.5\,GPa), $B_\mathrm{c2}$ is relatively small, i.e., $\sim$16\,T between 5\,K and 10\,K,\cite{Terashima_Eu1} and hence can be traced down to very low temperatures.
EuFe$_2$As$_2$ is unique in that the localized Eu$^{2+}$ moments exhibit an AF order below 20\,K\,\cite{Raffius_Mossbauer,Jeevan_single,Ren,Xiao_PRB09,Jiang_NJP09} in addition to an AF order arising from the FeAs layers at $T_\mathrm{\>0}$\,$\sim$190\,K.
$T_\mathrm{N}$ of the Eu$^{2+}$ moments is insensitive to pressure, and the AF order occurs in the $P$-induced superconducting state as evidenced by magnetic and heat capacity measurements under high-pressure.\cite{Terashima_Eu1,Miclea,Kurita_Eu1,Matsubayashi_Eu}
Despite the AF order, which is produced by a weak interlayer interaction, the dominant interaction among the Eu$^{2+}$ moments is the intralayer ferromagnetic (FM) interaction, and hence the FM alignment of the Eu$^{2+}$ moments is easily achieved by the application of 1\,$\sim$\,2\,T even below $T_\mathrm{N}$ at ambient pressure as well as under high pressure.\cite{Xiao_PRB09,Jiang_NJP09,Terashima_Eu1,Terashima_mag,Miclea,Xiao_PRB10}
Thus, EuFe$_2$As$_2$ provides an excellent opportunity where a long-standing issue of the interplay between superconductivity and magnetism can be studied in a high-$T_\mathrm{c}$ material using high-quality single crystals.

In this report, we present the $B_\mathrm{c2}$$-$$T_\mathrm{c}$ phase diagram of EuFe$_2$As$_2$ at a pressure of 2.5\,GPa and minimum temperature of 1.6\,K via high-field resistivity measurements up to 27\,T, and discuss the origin of the distinctive $B_\mathrm{c2}$ curves.


Single crystals of EuFe$_2$As$_{2}$ were prepared via the Bridgman method from a stoichiometric mixture of the constituent elements. 
The samples analyzed in this study were obtained from the same batch (residual resistivity ratio $RRR$\,=\,7) as that used in Refs.\,\onlinecite{Terashima_Eu1,Kurita_Eu1} and \onlinecite{Terashima_mag}.
The resistivity of two samples, denoted by $^{\#}$1 and $^{\#}$2, was simultaneously measured at $P$\,=\,2.5\,GPa via an ac four-probe method in a $^4$He cryostat ($T$\,$\ge$\,1.6\,K). 
Sample\,$^{\#}$1 ($^{\#}$2) was aligned with the $ab$-plane ($c$-axis) parallel to the longitudinal direction of a hybrid-type piston cylinder pressure cell\,\cite{PistonCell} for $B$\,$\parallel$\,$ab$ ($\parallel$\,$c$) measurements. 
For both samples, the magnetic field was applied along the piston cylinder axis in a direction perpendicular to that of the current.
To generate hydrostatic pressure, Daphne\,7474 (Idemitsu Kosan) oil, which remains in the liquid state up to 3.7\,GPa at room temperature,\cite{Daphne7474} was used as the pressure-transmitting medium. 
The samples were gradually cooled at an average rate of 0.5\,K/min. 
The pressure was calibrated at 4.2\,K by the resistance change of a Manganin wire.\cite{Terashima_Eu1} 
Magnetic fields up to 27\,T were produced by a water-cooled resistive magnet installed at the Tsukuba Magnet Laboratory, National Institute for Materials Science.
A 17-T superconducting magnet was used for preliminary resistivity studies. 
In this study, the magnetic field $B$ denotes an externally applied field, and the magnetization within a sample (up to $\sim$\,0.9\,T\,\cite{Terashima_mag}) is neglected. 


\begin{figure}
\begin{center}
\includegraphics[width=0.8\linewidth]{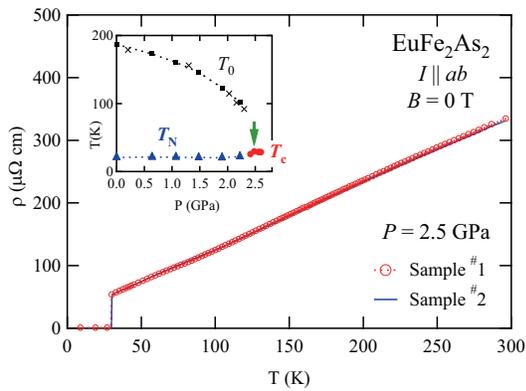}
\end{center}
\caption{(Color online) $\rho$ vs $T$ for EuFe$_2$As$_{2}$ at $P$\,=\,2.5\,GPa for samples\,$^{\#}$1 and $^{\#}$2 in the absence of applied field. 
The direction of current $I$ is $I$\,$\parallel$\,$ab$. 
The inset illustrates the $T$$-$$P$ phase diagram of EuFe$_2$As$_{2}$.\cite{Kurita_Eu2}
$T_\mathrm{\>0}$ and $T_\mathrm{N}$ denote the temperatures of the AF order arising from the FeAs layers and  localized Eu$^{2+}$ moments, respectively.
The solid circles denote $T_\mathrm{c}$ determined under the criterion $\rho$\,=\,0. 
The crosses denote the values obtained from Ref.\,\onlinecite{Miclea}.  } \label{fig1}
\end{figure}

Figure~\ref{fig1} shows the temperature dependence of the resistivity, $\rho(T)$, for the two samples, $^{\#}$1 and $^{\#}$2, at $P$\,=\,2.5\,GPa in the absence of an applied field. 
For both samples, $\rho$ exhibits virtually $T$-linear dependence in the broad temperature range above $T_\mathrm{c}$ without any anomaly due to the AF order of the FeAs layers.
This observation is consistent with the phase diagram shown in the inset:\cite{Miclea,Kurita_Eu2} $P$\,=\,2.5\,GPa is just above the critical pressure $P_\mathrm{c}$, where $T_\mathrm{\>0}$\,$\rightarrow$\,0, as indicated by the arrow.
Similar $\rho$\,$\sim$\,$T$ behavior was also reported in several optimally doped Fe-based SCs.\cite{Tlinear1,Tlinear2} 
However the reason for such behavior has not been verified thus far.
Both samples exhibit a sharp transition to zero resistivity at $T_\mathrm{c}$\,=\,30 K; the reentrant-like behavior as reported in Ref.\,\onlinecite{Miclea} is not observed for either sample at this pressure.
Our previous work\,\cite{Kurita_Eu1} indicates that reentrant-like behavior may be observed for $P$\,$<$\,$P_\mathrm{c}$ but not for $P$\,$>$\,$P_\mathrm{c}$ (as long as $P$ is not far from $P_\mathrm{c}$) in our single crystals.
Since both $T_\mathrm{c}$ and $B_\mathrm{c2}$ attain maximum values at $P$\,$\approx$\,$P_\mathrm{c}$, followed by a monotonic decrease with increasing $P$.\cite{Kurita_Eu2}, $B_\mathrm{c2}$ determined at 2.5\,GPa in this study is expected to be close to its maximum value.

\begin{figure}
\begin{center}
\includegraphics[width=0.8\linewidth]{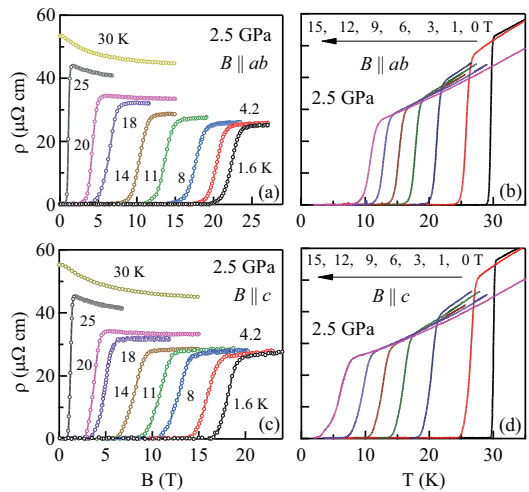}
\end{center}
\caption{(Color online) (a) $\rho$ vs $B$ and (b) $\rho$ vs $T$ for $B$\,$\parallel$\,$ab$ (sample\,$^{\#}$1) and  (c) $\rho$ vs $B$ and (d) $\rho$ vs $T$ for $B$\,$\parallel$\,$c$ (sample\,$^{\#}$2) in EuFe$_2$As$_{2}$ at $P$\,=\,2.5\,GPa.} \label{fig2}
\end{figure}

\begin{figure}
\begin{center}
\includegraphics[width=0.8\linewidth]{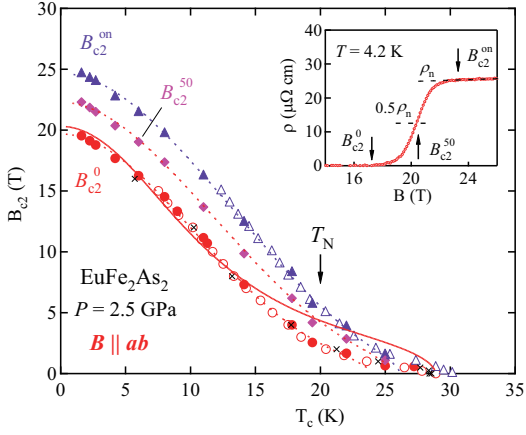}
\end{center}
\caption{(Color online) $B_\mathrm{c2}$$-$$T_\mathrm{c}$ phase diagram of EuFe$_2$As$_{2}$ for $B$\,$\parallel$\,$ab$ at 2.5\,GPa. 
The values of $B_\mathrm{c2}$ are determined under three different criteria, as illustrated for $\rho(B)$ data at 4.2\,K (inset). 
The solid or open symbols denote $B_\mathrm{c2}$ determined from $\rho(B)$ and $\rho(T)$ measurements, respectively. 
The solid and dashed curves are fits to Eq.(\ref{eq1}). 
{\bf $\times$} denotes the previous $B_\mathrm{c2}^\mathrm{\>0}$ result deduced from an ac-$\chi$ measurement for $B$\,$\parallel$\,$ab$\,\cite{Terashima_Eu1}. 
The arrow indicates $T_\mathrm{N}$ of Eu$^{2+}$ moments in the superconducting state in the absence of an applied field at 2.6\,GPa\,\cite{Terashima_Eu1}. } \label{fig3}
\end{figure}

\begin{figure}
\begin{center}
\includegraphics[width=0.8\linewidth]{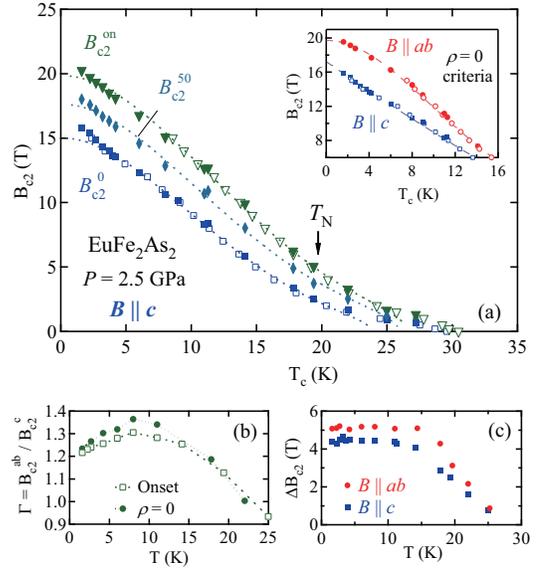}
\end{center}
\caption{(Color online) (a) $B_\mathrm{c2}$$-$$T_\mathrm{c}$ phase diagram of EuFe$_2$As$_{2}$ for $B$\,$\parallel$\,$c$ at 2.5\,GPa. 
The solid and open symbols denote $B_\mathrm{c2}$ deduced from $\rho(H)$ and $\rho(T)$ data, respectively.  
The dashed curves are fits to Eq.(\ref{eq1}). 
The inset shows $B_\mathrm{c2}^\mathrm{\>0}$ vs $T_\mathrm{c}$ for $B$\,$\parallel$\,$ab$ and $B$\,$\parallel$\,$c$. 
The dashed curves are fits for $T$\,=\,0 extrapolation (see text). 
(b) $T$-variation of an anisotropy parameter, $\Gamma$\,=\,$B_\mathrm{c2}^{ab}$/$B_\mathrm{c2}^{c}$, determined by the onset and zero-resistivity. 
(c) $T$-dependence of the superconducting transition width, $\Delta B_\mathrm{c2}$\,(=\,$B_\mathrm{c2}^\mathrm{on}$$-$$B_\mathrm{c2}^\mathrm{\>0}$), for $B$\,$\parallel$\,$ab$ and $B$\,$\parallel$\,$c$.
} \label{fig4}
\end{figure}

Figure~\ref{fig2}(a)-(d) shows the resistivity of EuFe$_2$As$_{2}$ at 2.5 GPa as a function of $B$ and $T$ for the two orientations $B$\,$\parallel$\,$ab$ and $B$\,$\parallel$\,$c$. 
A magnetic field of 27\,T is sufficient to recover the normal state at the minimum temperature, 1.6\,K ($\approx$\,0.05\,$\times$\,$T_\mathrm{c}$), for both orientations.
Using the data in Fig.~\ref{fig2}(a) and (b), the $B_\mathrm{c2}$$-$$T_\mathrm{c}$ phase diagram of EuFe$_2$As$_{2}$ is constructed for $B$\,$\parallel$\,$ab$ at 2.5\,GPa, as shown in Fig.~\ref{fig3}. 
Three sets\,$-$\,$B_\mathrm{c2}$, $B_\mathrm{c2}^\mathrm{on}$ (onset), and $B_\mathrm{c2}^x$ ($x$\,=\,0 and 50, $x$\,\% of the normal state resistivity $\rho_\mathrm{n}$)\,$-$\,are plotted, and their definitions are illustrated in the inset. 
The solid and open symbols are obtained from the $\rho(B)$ and $\rho(T)$ measurements, respectively. 
$B_\mathrm{c2}^\mathrm{\>0}$ is consistent with the previous result\,($\times$) obtained from the ac-$\chi$ measurement for $B$\,$\parallel$\,$ab$.\cite{Terashima_Eu1}
Note that all the curves of $B_\mathrm{c2}$ for $B$\,$\parallel$\,$ab$ ($B_\mathrm{c2}^\mathrm{ab}$) obtained under different criteria exhibit qualitatively similar $T$-dependence. 
The same is also true for $B_\mathrm{c2}$ for $B$\,$\parallel$\,$c$ ($B_\mathrm{c2}^\mathrm{c}$), as shown in Fig.~\ref{fig4}(a). 
$T_\mathrm{N}$ at zero field is indicated by an arrow in Figs.~\ref{fig3} and ~\ref{fig4}.  
However, we note that, since the AF order of the Eu$^{2+}$ moments is destroyed by an applied field of $\sim$\,1\,T,\cite{Terashima_Eu1} the $B_\mathrm{c2}$ curves for both $B$\,$\parallel$\,$ab$ and $B$\,$\parallel$\,$c$ are in the paramagnetic or field-induced FM state of the Eu$^{2+}$ moments.

A distinctive feature, the concave (upward) curvature of $B_\mathrm{c2}^\mathrm{ab}$ around 20\,K, has not been reported in other Fe-based SCs without localized magnetic ions. 
Therefore, it is likely related to the magnetic state of the Eu$^{2+}$ moments.
Similar concave $B_\mathrm{c2}(T)$ curves have been reported in Chevrel-phase compounds such as (Eu,$M$)Mo$_6$S$_8$ ($M$\,=\,Sn,\cite{JP_SnEuMoS} La,\cite{JP_LaEuMoS} etc.) and EuMo$_6$S$_8$ under pressure.\cite{JP_EuMoS}
In these systems, the conduction electrons are subjected to an exchange field $B_J$ in addition to an applied field via AF coupling with the Eu$^{2+}$ localized magnetic moments.
Note that the concave curvature is an indication of the negative sign of $B_J$; $B_J$ is antiparallel to the applied field.\cite{JP_effect, JP_EuMoS} 
Within a multiple pair-breaking picture, $B_{c2}$ in the dirty limit of three-dimensional SCs with negative $B_J$ can be expressed by\,\cite{WHH1966,JP_effect2}

\begin{equation}
\begin{aligned}
  \ln \frac{1}{t}   = & \biggl( \frac{1}{2}  + \frac{i\lambda_\mathrm{so}}{4\gamma} \biggr) \times \Psi \biggl( \frac{1}{2} + \frac{h+i\lambda_\mathrm{so}/2+i\gamma}{2t} \biggr) \\
  +  \biggl(   \frac{1}{2} & - \frac{i\lambda_\mathrm{so}}{4\gamma} \biggr)  \times \Psi \biggl( \frac{1}{2} + \frac{h+i\lambda_\mathrm{so}/2-i\gamma}{2t} \biggr)  -  \Psi  \biggl( \frac{1}{2}  \biggr)\\
 \gamma=&[\alpha^2  (h+ h_J)^2-\lambda_\mathrm{so}^2]^{\frac{1}{2}} \label{eq1}
\end{aligned}
\end{equation}
where $\Psi$ and $\lambda_\mathrm{so}$ are the digamma function and spin-orbit scattering parameter, respectively.
The magnetic scattering parameter $\lambda_\mathrm{m}$ used in the complete formula\,\cite{WHH1966,JP_effect2} is typically ignored for simplicity.\cite{JP_EuMoS,JP_effect3} 
The Maki parameter $\alpha$ is defined as $\sqrt{2}$\,$B_\mathrm{c2}^{*}$/$B_\mathrm{p}$, using the orbital critical field $B_\mathrm{c2}^*$ at $T$\,=\,0 and the Pauli-Clogston paramagnetic limit $B_\mathrm{p}$\,\cite{CClimit1}.
Reduced units\,$-$\,$t$\,=\,$T$/$T_\mathrm{c}$, $h$\,=\,0.281\,$B_\mathrm{c2}$/$B_\mathrm{c2}^{*}$, and $h_J$\,=\,0.281\,$B_J$/$B_\mathrm{c2}^{*}$\,$-$\,are employed.
We assume $B_J$\,=\,$\beta$\,$M$ ($\beta$:\,constant), where the magnetization $M$ is modeled within a molecular-field approximation.\cite{Magnetization}
To simplify the following discussions, $\alpha$ for $B$\,$\parallel$\,$ab$ is set to 3, a typical value for ``122" systems.

The solid curve in Fig.~\ref{fig3} was calculated from Eq.(\ref{eq1}) for $B_\mathrm{c2}^\mathrm{\>0}$ data with $T_\mathrm{c}$ set to the experimental value $T_\mathrm{c}$\,=\,29\,K.
The fit yields a parameter set ($\lambda_\mathrm{so}$, $\beta$)\,=\,(7.9, $-$187).
$\beta$\,=\,$-$187 indicates that the maximum of $|B_J|$, $B_J^\mathrm{m}$, is around 168\,T.
The fit captures the qualitative characteristics of the experimental $B_\mathrm{c2}^\mathrm{\>0}$ curve satisfactorily, especially the positive curvature below $T_\mathrm{N}$\,=\,20\,K, and shows that the low value of $B_\mathrm{c2}$ (compared to other Fe-based SCs' with similar $T_\mathrm{c}$ values) is due to the large $B_J$, which is a consequence of a large Eu$^{2+}$ magnetization due to the field-induced FM alignment of the Eu$^{2+}$ moments.
However, its deviation from the experimental curve is also noticeable at low fields near $T_\mathrm{c}$.
This disagreement probably indicates that the phase diagram in this $T$-range is affected by a subtle competition between superconductivity and magnetic fluctuations, and it is beyond the scope of Eq.(\ref{eq1}), which assumes a homogeneous $B_J$ produced by paramagnetic spins.
Since the dominant interaction among the Eu$^{2+}$ moments is the intralayer FM interaction,\cite{Jiang_NJP09,Terashima_mag,Xiao_PRB09,Xiao_PRB10} the FM fluctuations develop when $T$ is lowered to $T_\mathrm{N}$, as evidenced by the enhancement of the magnetic susceptibility as $T$\,$\rightarrow$\,$T_\mathrm{N}$.\cite{Jiang_NJP09,Terashima_Eu1,Terashima_mag}
Such FM fluctuations may be detrimental to superconductivity.
One way to phenomenologically overcome this problem and to improve the fit in a $T$-range not close to $T_\mathrm{c}$ is to use a reduced value of $T_\mathrm{c}$.
Thus, the three dotted curves are calculated using the reduced $T_\mathrm{c}$ value.
They reproduce the experimental curves excellently over the entire $T$-range, with a minimum temperature of 1.6\,K.
For $B_\mathrm{c2}^\mathrm{\>0}$, we assumed ($\lambda_\mathrm{so}$, $\beta$, $T_\mathrm{c}$)\,=\,(2.7, $-$83, 24\,K), where $B_J^\mathrm{m}$\,$\sim$\,75\,T.
Here, it may be worthwhile to compare the parameters with those of the Chevrel compounds.
The comparison revealed that the obtained $\lambda_\mathrm{so}$ is comparable to that found in the Chevrel-type Eu compounds,\cite{JP_EuMoS,JP_effect3}
and $B_J^\mathrm{m}$ in EuFe$_2$As$_2$ is a few times greater than that reported in the Chevrel-type Eu compounds.\cite{JP_EuMoS,JP_effect3}
We note that the concave curvature of $B_\mathrm{c2}$ in EuFe$_2$As$_2$ essentially differs from the positive curvatures often observed in highly two-dimensional SCs such as high-$T_\mathrm{c}$ cuprates.
In the latter, the curvature is highly dependent on what criterion is chosen to define $B_\mathrm{c2}$, and it is most likely affected by the vortex lattice phase transitions, i.e. from a vortex-liquid state to a vortex-solid state.\cite{Ando_Bc2}

Figure~\ref{fig4}(a) shows the $B_\mathrm{c2}$$-$$T_\mathrm{c}$ phase diagram of EuFe$_2$As$_{2}$ for $B$\,$\parallel$\,$c$ at 2.5\,GPa,\cite{DiscrepancyinBc2} determined in the same manner as that used for $B_\mathrm{c2}^{ab}$.
A concave curvature around 20\,K is also visible for the $B_\mathrm{c2}^\mathrm{c}$ curves.
The dashed curves are calculated using the parameters comparable to those used for $B_\mathrm{c2}^{ab}$, 
i.e., for $B_\mathrm{c2}^{\>0}$, the fit gives ($\alpha$, $\lambda_\mathrm{so}$)\,=\,(1.9, 2.6) when we assume ($\beta$, $T_\mathrm{c}$)\,=\,($-$83, 24\,K), identical to the values used for $B_\mathrm{c2}^{ab}$. 
The calculated curves tend to saturate below 3\,K, whereas the experimental curves appear to increase linearly as $T$ decreases to zero.
The unsaturation of $B_\mathrm{c2}^{c}$ has been observed in other Fe-based SCs,\,\cite{twobandLaFeAsO1,Yuan_BaK122, twobandSrFeCo2As2, Kano} and it has been explained using a two-band model.
Figure~\ref{fig4}(b) shows the anisotropy ratio, $\Gamma$\,=\,$B_\mathrm{c2}^{ab}$\,/\,$B_\mathrm{c2}^{c}$, calculated from $B_\mathrm{c2}^\mathrm{\>0}(T)$ and $B_\mathrm{c2}^\mathrm{on}(T)$.
In spite of the quasi-two-dimensional layered structure in EuFe$_2$As$_{2}$, we obtain a small value of $\Gamma$, ranging between 0.9 and 1.4, which is comparable to that obtained for other ``122" compounds.\cite{twobandSrFeCo2As2,Yuan_BaK122, Kano,BaFeCo2As2_Yamamoto}
In contrast to the monotonic decrease in $\Gamma$ with decreasing $T$ in other ``122" compounds, $\Gamma$ in EuFe$_2$As$_{2}$ exhibits a broad maximum at around 8\,K, which is likely ascribed to the presence of the $B_J$.

In order to compare the magnitude of $B_\mathrm{c2}$(0) with that of other Fe-based SCs, we estimate it by extrapolating the low-$T$ data to $T$\,$=$\,0, as shown by the dashed curves in the inset of  Fig.~\ref{fig4}(a).  
For the extrapolations, an empirical expression, $B_\mathrm{c2}(t)=B_\mathrm{c2}(0)$(1\,$-$\,$t^2$)/(1\,$+$\,$t^2$),\cite{Extraporation} and a linear fit are used for $B_\mathrm{c2}^\mathrm{ab}$ and $B_\mathrm{c2}^\mathrm{c}$, respectively. 
We obtain $B_\mathrm{c2}^\mathrm{ab}(0)$\,=\,24.7\,T and 19.7\,T and $B_\mathrm{c2}^\mathrm{c}$(0)\,=\,21.5\,T and 17.2\,T for $B_\mathrm{c2}^\mathrm{on}$ and $B_\mathrm{c2}^\mathrm{\>0}$, respectively. 
$B_\mathrm{c2}(0)$ in EuFe$_2$As$_2$ is significantly lower than $B_\mathrm{c2}(0)$\,$>$\,50\,T in other Fe-based SCs at $T_\mathrm{c}$\,=\,20\,-\,30\,K.\cite{twobandLaFeAsO1,Yuan_BaK122, twobandSrFeCo2As2, Kano}
The width of the superconducting transition, $\Delta B$ (=\,$B_\mathrm{c2}^\mathrm{on}$\,$-$\,$B_\mathrm{c2}^\mathrm{\>0}$), increases as $T$ decreases to 15\,K for both $B$\,$\parallel$\,$ab$ and $B$\,$\parallel$\,$c$ [Fig.~\ref{fig4}(c)].
Below 15\,K, $\Delta B$ is virtually $T$-independent, as reflected by the parallel-shifts of the $\rho(B)$ curves in Fig.~\ref{fig2}(a) and (c).
The $T$-dependence may correlate with the development of $M$; $M$ at $B$\,=\,$B_\mathrm{c2}(T)$ increases rapidly as $T$ decreases from $T_\mathrm{c}$, but it is virtually saturated below $\sim$\,15\,K.\cite{Jiang_NJP09}
At 1.6\,K, $\Delta B$ is estimated as 5.1\,T and 4.4\,T for $B$\,$\parallel$\,$ab$ and $B$\,$\parallel$\,$c$, respectively.
The relatively narrow transition width at low $T$, which is also observed in Ba(Fe,Co)$_2$As$_2$,\cite{BaFeCo2As2_Yamamoto,Kano} signifies a strong vortex pinning force in EuFe$_2$As$_{2}$.


In conclusion, we carried out high-field resistivity measurements up to 27\,T for EuFe$_2$As$_2$ at 2.5\,GPa, and we constructed the $B_\mathrm{c2}$$-$$T_\mathrm{c}$ phase diagram down to a minimum temperature of 1.6\,K.
Our analysis was based on a multiple pair-breaking model, and it revealed that the distinctive $B_\mathrm{c2}$ curves with positive curvature and the reduced $B_\mathrm{c2}$ values can be attributed to the substantial negative exchange field from the Eu$^{2+}$ moments. 
The low temperature anisotropy at 1.6\,K, $\Gamma$\,=\,1.2, is comparable to the results obtained for other ``122" systems.


\begin{thebibliography}{25}
\bibitem{Magnetization}
The model reproduces experimental $c$-axis magnetization satisfactorily.\cite{Terashima_mag}  
The anisotropy of magnetization is small, except for a low-$T$ and low-$B$ region;\cite{Jiang_NJP09} hence, we use the same model for both $B$\,$\parallel$\,$ab$ and $B$\,$\parallel$\,$c$.
\bibitem{DiscrepancyinBc2}
There is discrepancy between the present and previous $B_\mathrm{c2}^\mathrm{c}$ data.  
The latter was determined from ac-$\chi$.\cite{Terashima_Eu1}  
This could be due to the difference in the applied pressure and/or sample variations. 
 In this study, $\rho$ was simultaneously measured using the two samples from the same piece.
\end{thebibliography}
\end{document}